\documentclass{article}
\usepackage[whole]{bxcjkjatype}
\usepackage{graphicx}
\usepackage[fleqn]{amsmath}
\usepackage{amssymb}
\usepackage{url}
\usepackage[unicode]{hyperref}

\begin{document}
\title{団代数と超対称ゲージ理論\thanks{雑誌数理科学2015年3月号特集「団代数をめぐって：新たな共通構造の認識」}}
\date{}

\author{山崎 雅人（東京大学カブリ数物連携宇宙研究機構）}
\maketitle

\section{団代数と物理学者の邂逅}

筆者は理論物理学者であるが，
「数物連携」と名のつく研究所に所属しており
数学者とも交流が深い．
理論物理学者と純粋数学者とでは
研究の手法も動機も大きく異なるが，
昨今の素粒子理論の研究においては，
純粋数学において育てられた概念に，思わぬところで
出くわすことも少なくない．

本特集で取り上げられている\textbf{団代数}（クラスター代数）
もその一つのめざましい例である．本稿では，
超対称場を研究する物理学者たちが，
いかにして団代数に出会ったかを
説明したい．

\section{電場と磁場のハーモニー}

我々が考えたいのは我々の住む4次元時空，
すなわち空間3次元，時間１次元において定義された
\textbf{ゲージ場の理論}である．

ゲージ理論はゲージ群$G$を指定することによって得られるのであった．
簡単な例として，ゲージ群が$U(1)$の
場合を考えよう．これはおなじみの電磁気学であり，
$U(1)$ゲージ場は光子を表す．
電磁気学では，光子だけではなく，
光子と相互作用する物質場，例えば電子を
考える．これはゲージ群のもとで，とある
電荷$e$をもった場のことである．

電磁場は磁場と電場からなっていたが，電子は電場のみに対して
チャージを持っていた．逆に，磁場のみに対してチャージ（これを磁荷$g$と書こう）を持つ
のが\textbf{磁気モノポール}（磁気単極子）である．
ディラックが気づいたのは，モノポールと電子の両者を考えることで，
マックスウェル方程式は電場と磁場の入れ替えについて対称な形になることであった．さらに彼は，
電荷と磁荷の満たすべき量子化条件
\begin{align}
e g \in 2\pi \hbar\, \mathbb{Z}
\end{align}
を導いたのであった\footnote{これらのことについては，例えば数理科学誌2014年7月号の特集
「モノポールの謎」を参照されたい．}．

ここまでは電荷ないし磁荷のみを持つ理論を考えたが，
両方を同時に持つ粒子（\textbf{ダイオン}）を考えてもよい．その場合の量子化条件は，
二つの粒子の電荷・磁荷の組（以下単にチャージと呼ぶ）を
$\gamma_{i=1,2}=(e_i, g_i)$としたとき，その
ペアリング$\langle \gamma_1 , \gamma_2 \rangle$の量子化条件として表される：
\begin{align}
\langle \gamma_1, \gamma_2 \rangle := e_1 g_2- e_2 g_1\in 2\pi \hbar\, \mathbb{Z}\ .\label{pairing}
\end{align}
すぐにわかるように，このペアリングは
完全半対称$\langle  \gamma_1 , \gamma_2 \rangle = -\langle \gamma_2 , \gamma_1 \rangle$である．
ゲージ群をより一般の可換ゲージ群$U(1)^r$にした時も，チャージ$e, g$がそれぞれ$r$成分を持つ
ベクトル$\vec{e}, \vec{g}$になることを除けば量子化条件は同様である．
\eqref{pairing}式で導入したチャージのペアリングは後で見るように
箙の定義に用いるので重要である．

\section{4次元$\mathcal{N}=2$理論のクーロンブランチ}
ここまでの議論では
例えばゲージ場のラグランジアンの
具体的な形を直接必要としたわけではない．
従って，もともとゲージ群が非可換の設定から出発しても，最終的にゲージ群がその可換部分に破れている限り，議論は同様であると期待される．

ゲージ群が破れると聞いてまず思い出すのはヒッグズ効果である：例えば$SU(3)$ゲージ群の基本表現に属する
クオークが期待値を持てばゲージ群は破れる．しかし，この場合ゲージ群は一般には完全に破れてしまい，可換ゲージ群すら残らない．

そこで，通常のクオークの代わりにゲージ群に対して随伴表現で変換する場$\sigma$（つまりゲージ群の元$g$のもと$\sigma\to g^{-1} \sigma g$と変換する場）が存在し，その場が（一般の）期待値を持ったとしよう．このとき，ゲージ群の非可換部分は破れるが，ゲージ群の可換部分群$U(1)^{r}$が全て残ることになる（随伴表現はゲージ群の可換部分群に対して変換を受けない）．ここで$r$はゲージ群のランクとよばれる量である．

このような状況のうち典型的なものとして，\textbf{4次元$\mathcal{N}=2$理論}を考える．
超対称性はボソンとフェルミンを入れ替える対称性であるが，$\mathcal{N}=2$対称性では
独立な超対称変換が二つ存在するので，その二つを組み合わせるとボソンが別のボソンに移されることになる．
特に，ボソンであるゲージ場$A_{\mu}$に別のボソンのペアが存在し，それが先に述べたスカラー
場$\sigma$である．ゲージ場はゲージ群の随伴表現で変換するので，それのペアである
$\sigma$も期待通り随伴表現で変換する．
$\mathcal{N}=2$対称性を持つ理論では，$\sigma$が期待値を持つ真空が存在し，そこでは可換なゲージ場が残るので\textbf{クーロンブランチ}（クーロン枝）\footnote{枝と呼ばれるのは特定の点（例えば共形不変性を持つ点）から複数の「枝」が伸びてくるからなのではないかと想像する．}と呼ばれる．我々はこの真空における4次元$\mathcal{N}=2$理論に対する低エネルギーでの振る舞いを調べることにしよう．そのような低エネルギー有効理論を解くのが，90年代半ばに現れたザイバーグ・ウィッテン理論
であり，これまで本誌上でもたびたび取り上げられてきた．団代数は，この古典的な設定とその一般化を議論する中で現れてきたのである．

\section{BPS粒子から箙へ}

クーロンブランチではゲージ群は可換であるが，
我々の出発点は非可換ゲージ群をもつゲージ理論である．
物質場の数が多すぎない時，理論は漸近自由性を持ち
我々の興味のある低エネルギーでは強結合領域にありその
直接の解析は容易ではない．

ここでは，理論そのものを直接調べる代わりに，理論のスペクトラムを調べることにしよう．
つまり，どういうチャージを持った安定な粒子が，幾つ存在するかを考えるのである．
$\mathcal{N}=2$超対称性からの制限を最大限に活用するために，特に$\mathcal{N}=2$超対称性のうちの
最大限（今の場合は半分の）超対称性を保つ粒子のみを考えることにしよう．
このような粒子のことを
\textbf{BPS粒子}（今の場合は正確には$\frac{1}{2}$-BPS粒子）と呼ぶ．

BPS粒子の特徴として，チャージ$\gamma$を持つBPS粒子の質量は，そのチャージから
定まる複素数$Z_{\gamma}$（\textbf{セントラルチャージ}）の絶対値$\big| Z_{\gamma}  \big|$で決まる．
一方，$Z_{\gamma}$の位相部分は，$\mathcal{N}=2$対称性のうちどの$\mathcal{N}=1$超対称性
を保つかを指定している．さらに，$Z_{\gamma}$は，$\gamma$に対して線形である：
$Z_{\gamma_{1}+\gamma_2}=Z_{\gamma_1}+Z_{\gamma_2}, Z_{n \gamma}=n Z_{\gamma}$.

クーロンブランチでは破れずに残った可換ゲージ群$U(1)^r$が存在するので，
粒子はそのゲージ群についてのチャージ$\gamma\in \Gamma$を持つ．
ここで$\Gamma$は許されるチャージの全体であり，完全反対称ペアリングを持ちそのもとで
量子化条件\eqref{pairing}式を満たす．

我々は相対論的な局所場の理論から
出発しているので，CPT定理が成立し，特に
チャージ$\gamma$を持つ粒子が存在すれば，
逆のチャージ$-\gamma$を持つ反粒子も存在しなけれればならない．
したがって，粒子を数え上げる時には二つのうちどちらか一方だけを考えてやればよく，
$\Gamma$は二つの交わりを持たない和に分解する：
\begin{align}
\Gamma=\Gamma_{\rm 正} \cup \Gamma_{\rm 負} \ .
\end{align}

\begin{figure}
\centering\includegraphics[scale=0.3]{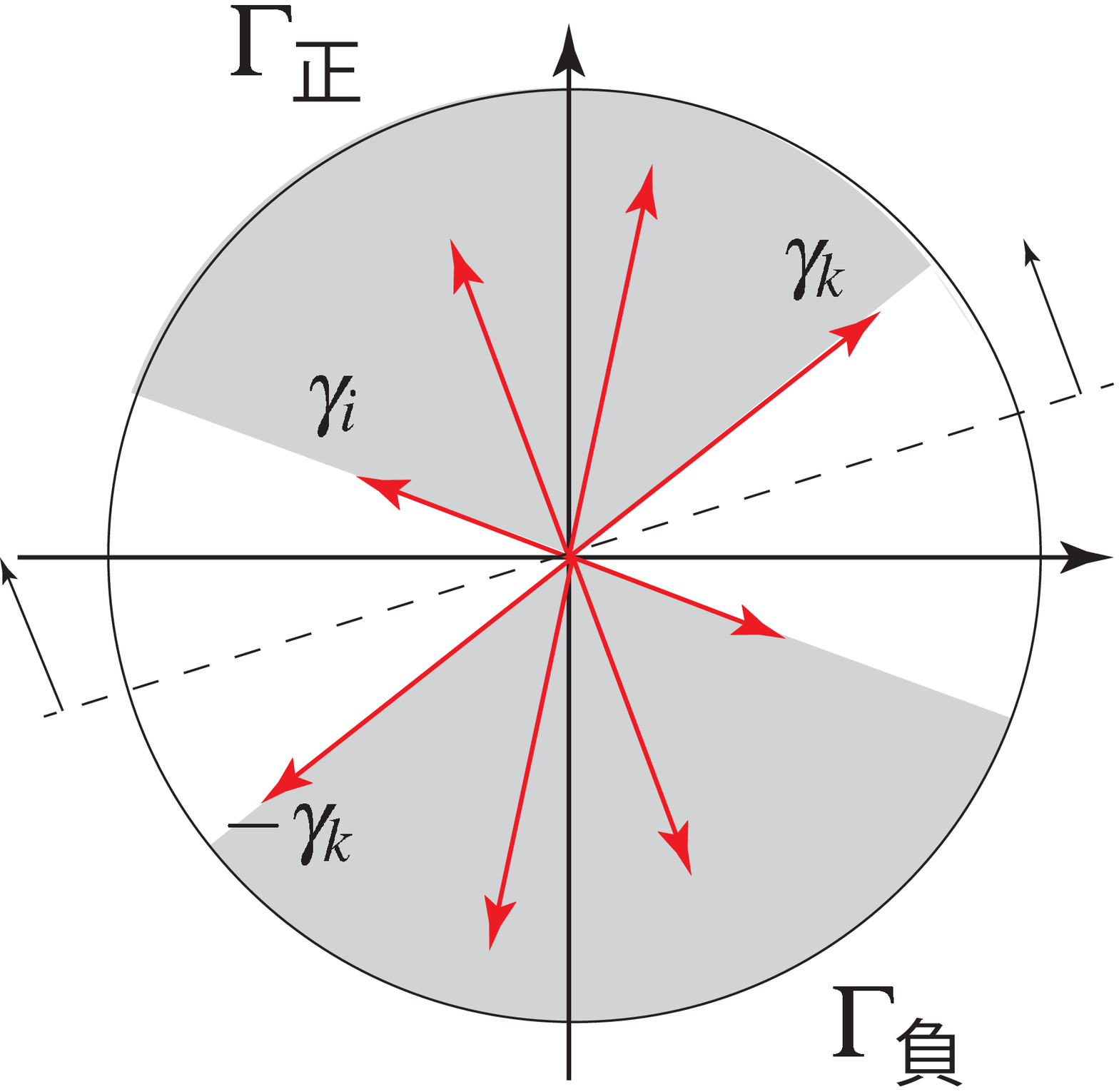}
\caption{チャージのなす集合$\Gamma$を$Z_{\gamma}$の偏角に応じて
正負に分割する．}
\label{half_plane}
\end{figure}

もっとも，この分解は一意ではない．
我々の目的のためには，
とある偏角$\zeta$を定めて先に導入した複素数$Z_{\gamma}$が複素平面上で
偏角が$[\zeta, \zeta+\pi]$にあるものを集めて$\Gamma_{\rm 正}$とすればよい（図\ref{half_plane}）：
\begin{align}
\gamma\in \Gamma_{\rm 正} \longleftrightarrow \textrm{Im} (e^{- i\zeta} Z(\gamma))>0 \ .
\label{positive}
\end{align}

さて，ここから箙を定義しよう．
$\Gamma_{\rm 正}$を生成する基底を$\{ \gamma_i \}$としよう：\footnote{
このような基底が存在するかどうかは明らかではない．例えば$\mathcal{N}=4$理論はそのような有限な基底を持たない．}
\begin{align}
\Gamma_{\rm 正} = \displaystyle\bigoplus_i \mathbb{Z}_{\ge 0} \gamma_i \ . \label{positive_span}
\end{align}
この時，
\begin{align}
b_{ij}:=\langle \gamma_i, \gamma_j \rangle 
\label{bdef}
\end{align}
により半対称行列$B=\left(b_{i,j} \right)$を，従って箙を定義するのである（本特集中西氏の
記事参照
）．この箙を\textbf{BPS箙}（えびら）と呼ぶことにしよう．
既に述べたようにペアリングは反対称であったから，$b_{ij}$も
反対称行列を与えることに注意されたい．

\eqref{bdef}式の定義は天下りだが，その背後にはちゃんと物理的意味がある．
BPS粒子は4次元$\mathcal{N}=2$理論の超対称ゲージ理論の中に存在する粒子であるが，それ自体
半分の超対称性を保っている．したがって，その粒子の上にいる人の立場に立てば，
4次元$\mathcal{N}=1$を持つ超対称量子力学が現れたように見える（粒子に対する場の理論は量子力学である）．
先に定義した箙はこの超対称量子力学（\textbf{箙超対称力学}）の定義データを与えるのである．

具体的には次のようにすれば良い：箙の頂点にはチャージ上で定義した基底の一つの元$\gamma_i$が対応する．
$\gamma=\sum_i n_i \gamma_i$なるチャージを考えた時には
$i$番目の頂点にはゲージ群$U(n_i)$を考え，二つの頂点$i$からでて$j$までを結ぶ辺には，
$U(n_i) \times U(n_j)$のもとで$(n_i, \bar{n}_j)$として変換する場を対応させる\footnote{正確にはスーパーポテンシャルにより相互作用も指定する必要がある．}\footnote{超対称量子力学の真空のモジュライ空間は数学的にはポテンシャル付き箙の
安定な表現のなすモジュライ空間であり，木村氏の解説に現れる箙多様体と近い関係にある．
}．

\section{箙の変異}

\begin{figure}
\centering\includegraphics[scale=0.3]{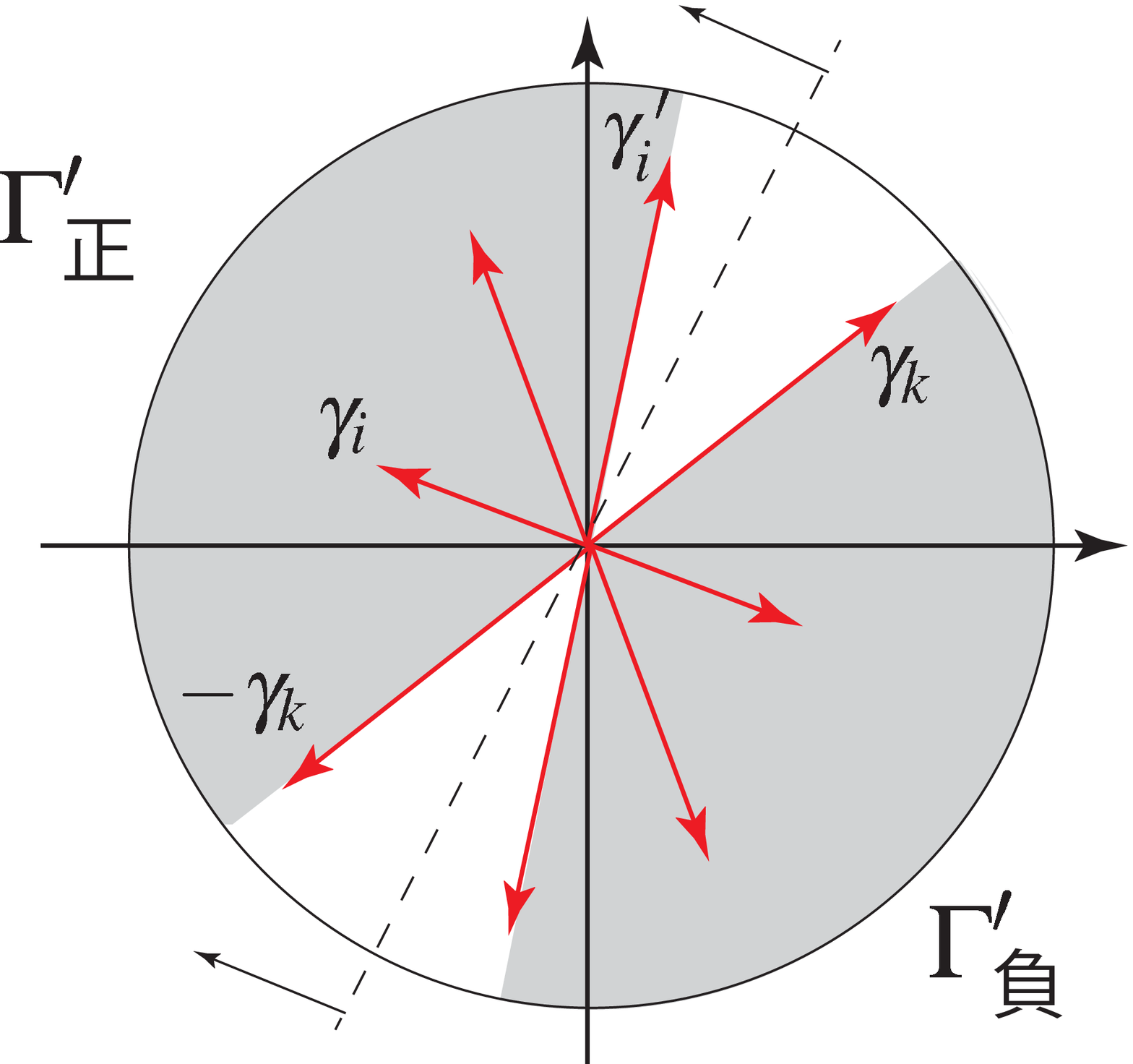}
\caption{$\zeta$の値を変化させることでチャージの正負への分割を変更する．
この時，$\Gamma_{\rm 正}$の基底も取り替える必要がある．}
\label{half_plane_change}
\end{figure}

さてここまでで箙とその物理的意味を説明してきたが，
ここまでの説明には不満足な点がある．それは，
正と負のチャージへの分割が一意ではないということだ．
このことは\eqref{positive}式が偏角$\zeta$に依存することからも明らかである．
特に，変化が起こるのは$\zeta$が基底のある元$\gamma_k$の偏角を越えて変化すると，
$\Gamma_{\rm 正}$が，従って箙が変化する（図\ref{half_plane_change}）．

それまで正のチャージを持っていた$\gamma_k$が
負のチャージを持つようになったとしよう．
このとき，明らかに$-\gamma_k$を新たな基底に加えなければならない:
\begin{align}
\gamma'_k=-\gamma_k \ .
\label{gchange_1}
\end{align}
しかし，\eqref{positive_span}を保つ必要があるので
他の基底の元も取り替える必要がある．その結果は
\begin{align}
\gamma_i \to \gamma_i + \left[b_{ik} \right]_{+}  \gamma_k
\label{gchange_2}
\end{align}
であることが知られている．
ただし，ここで$[x]_{+}:=\textrm{max}(x,0)$．
このとき，箙は\eqref{bdef}の定義から
\begin{align}
\begin{split}
&b_{ik} \to -b_{ik} \ , \\
&b_{ij} \to b_{ij} + \left[b_{ik} \right]_{+}  b_{kj}+ \left[b_{jk}\right]_{+} b_{ik}
\end{split}
\label{eq.mutate_rule}
\end{align}
と変化する．これは中西氏の解説中(5)式で導入された箙の変異
$B\to B'=\mu_k(B)$に他ならない．
こうして我々は箙の変異に辿りついた．

\eqref{gchange_2}の証明はここでは行わないが，例えば箙の表現論の立場から\cite{Alim:2011kw}のsection 3.1.2を参照されたい．また，より直接の説明として，先に説明した箙量子力学の双対性を用いたものがある．同じ箙からは2次元超対称場の理論も定義することができ，箙の変異はその双対性を表している\cite{Benini:2014mia}．それを1次元量子力学にに次元還元したのがここでの箙の変異を表すのだ．

\section{クラスター$y$変数とループ演算子}

クラスター代数はクラスター$x$変数や\textbf{$y$変数}（係数）と呼ばれる変数が重要な役割を果たした．
実は，我々の設定ではクラスター$y$変数はクーロンブランチの座標として現れるのである．

ここでは一般の4次元$\mathcal{N}=2$理論を考える代わりに，
$A_{N-1}$型の6次元$(2,0)$理論を点付き（つまり，穴のある）リーマン面$C$の上にコンパクト化した理論を考えることにしよう．
\begin{align}
\textrm{6次元理論} : \mathbb{R}^4\times C \longrightarrow
&
\, 
\textrm{4次元理論} : \mathbb{R}^4 \ .
\label{6d4d}
\end{align}
ここで，6次元$A_{N-1}$型$(2,0)$理論はその正体が明らかでない謎の理論であり直接の役には立たないが，
その$S^1$コンパクト化が5次元$\mathcal{N}=2$ $SU(N)$ゲージ理論を与えることはよくわかっている．
そこで，$\mathbb{R}^4$のうち一方向を$S^1$にコンパクト化することを考えよう：
\begin{align}
\begin{split}
& \textrm{6次元$A_N$型理論} : \mathbb{R}^3\times S^1 \times C \\
& \longrightarrow  \, 
\textrm{5次元$SU(N)$ゲージ理論} : \mathbb{R}^3 \times C \ .
\end{split}
\label{5d_compact}
\end{align}
となり，それは実行できる．\eqref{5d_compact}では先に$S^1$コンパクト化したが，順序を変えて
先に$C$にコンパクト化することすると，\eqref{6d4d}の
4次元理論が3次元理論に
$S^1$コンパクト化されることになる：
\begin{align}
\textrm{4次元理論} : \mathbb{R}^3\times S^1 \longrightarrow
\textrm{3次元理論} : \mathbb{R}^3 \ .
\label{4d3d}
\end{align}

さて，5次元理論のラグランジアンを用いて$C$上のBPS方程式を解析すると，
リーマン面$C$上に\textbf{ヒッチン・モジュライ}と呼ばれるモジュライ空間が現れる．
この空間は複素構造の取り方によって幾つかの記述があるが\footnote{
ヒッチン系はハイパーケーラ多様体であり$\mathbb{P}^1$でパラメーター付けされる
複素構造を持つ．}，
そのうちの一つは，$PSL(N, \mathbb{C})$平坦接続の空間であるというものである．
つまり，$C$上の複素接続$\mathcal{A}$で，
\begin{align}
\mathcal{F}=d \mathcal{A} + \mathcal{A} \wedge \mathcal{A}=0
\end{align}
を満たすものの全体を，ゲージ変換で割ったものである\footnote{
正確にはこのほかに$\mathcal{A}$と交換するスカラー場が存在し，それらが
クーロンブランチ以外の真空のブランチを記述している\cite{XieYonekura}．
}\footnote{但し，$C$の穴では$\mathcal{A}$のホロノミーを指定する境界条件が課される．}．
こうして現れた$PSL(N, \mathbb{C})$平坦接続の空間には，
自然な座標（\textbf{フォック・（ゴンチャロフ）座標}）\cite{FG}が
存在することが知られている．

\begin{figure}
\centering\includegraphics[scale=0.35]{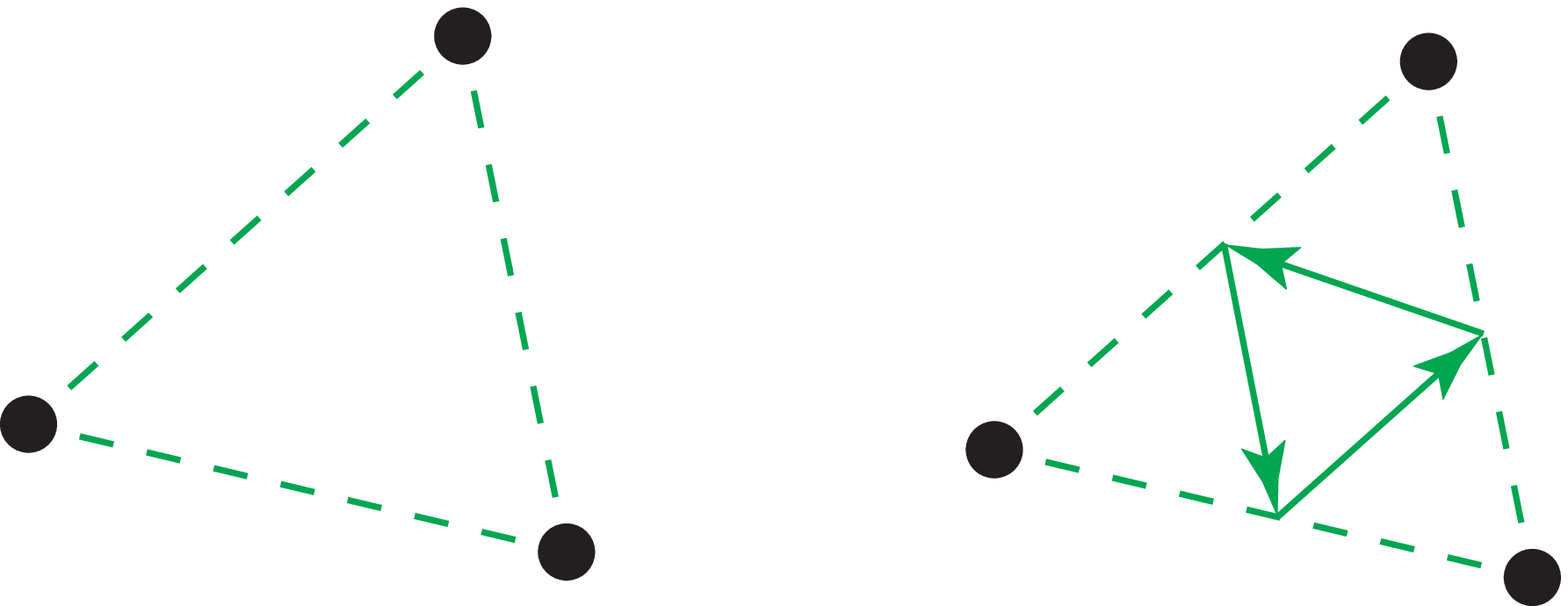}
\caption{理想三角形分割と，それから定めるBPS箙．}
\label{trig}
\end{figure}

ここでは簡単のため$N=2$の場合を考えることにしよう．
頂点を$C$の穴にもつ
三角形分割（そのような三角形分割を\textbf{理想三角形分割}と呼ぶ）を考え，
各三角形ごとに
図\ref{trig}の箙を書くことで$C$上に書かれた箙が$B=\left(b_{ij} \right)$が得られる（但し，$C$は十分な数の穴を持ち，そのような三角形分割が存在すると仮定する）．
このとき，箙の辺$i$（理論三角形分割の辺）複素変数
$y_{\gamma_i}$を対応させるとそれがヒッチン・モジュライの座標となり，
そのもとでのシンプレクティック形式は簡単な形
\begin{align}
\big\{ y_{\gamma_i}, y_{\gamma_j} \big\}= b_{i,j} 
\label{omega}
\end{align}
で与えられるというのが数学的な結果である．
なお，以下$y_{\gamma_1+ \gamma_2}:=y_{\gamma_1}+ y_{\gamma_2}$ $y_{n\gamma}=n y_{\gamma})$と定義することで，　一般の$\gamma\in \Gamma$に対し$y_{\gamma}$を定義しておくと便利である．

\eqref{omega}式を古典力学でおなじみのポアソン括弧とみなすことにすれば，
$y_{\gamma_i}$は有限次元の古典力学の相空間の座標に他ならない．
\eqref{pairing}と比較すれば，辺$i$は正チャージの基底$\gamma_i$に対応さて，
図\ref{trig}から定まる箙は4次元$\mathcal{N}=2$理論のBPS箙と同一視
するのが自然である．

さらに，座標$y_{\gamma}$はチャージ$\gamma$によって指定される
赤外理論での\textbf{ループ演算子}の期待値と同定される\cite{GMN}．
ここでループ演算子とは，一次元的に広がった演算子のことであり，その代表例は
ウィルソンラインであり，ゲージ場を閉路に沿って積分したものである：
4次元理論でのループ演算子は，4次元理論を$S^1$上で次元還元\eqref{4d3d}した時，
その$S^1$方向に巻きついているとすると3次元理論の粒子になる．
4次元でのクーロンブランチは随伴表現に値を持つ場$\sigma$の期待値によってパラメーター付けされていたのであった．3次元に次元還元すると，ゲージ場の$S^1$方向のゲージ場の積分は
3次元に新たな複素スカラー場$\bar{\sigma}$を与え，$\sigma, \bar{\sigma}$がペアをなして
3次元のクーロンブランチをパラメーター付けする．
したがって，ループ演算子の期待値の複素化
がクーロンブランチを指定する座標を与えるというのは自然であると納得できる．

厳密に3次元を考えるのではなく，$S^1$の半径$R$を
有限に保った時には
何が起こるだろうか？
\eqref{5d_compact}にたちもどると，
これは，5次元理論から6次元理論への持ち上げである．
超弦理論の言葉では，これはIIA型超弦理論の5次元のブレーン（D4ブレーン）が
M理論の6次元のブレーン（M5ブレーン）へ持ち上がる過程である．

このとき，3次元での粒子は4次元のループ演算子に持ち上がる．
粒子とは異なり，一般の電荷・磁荷を持つ
ループ演算子はお互いに交換しない演算子になることが知られている\cite{Polyakov,GMN}から，
座標$y_{\gamma_i}$は交換しない演算子$\hat{y}_{\gamma_i}$に置き換わるはずである．
自然な量子化は有限次元相空間\eqref{omega}のポアソン括弧を
演算子の交換関係に置き換えて得られる：
\begin{align}
\left[ \hat{y}_{\gamma_i}, \hat{y}_{\gamma_j} \right]= i \hbar \, b_{i,j}  \ .
\label{omega_q}
\end{align}
あるいは，$\hat{Y}_{\gamma}=e^{\hat{y}_{\gamma}}$で定義される変数を用いることにすると，
いわゆる\textbf{量子トーラス}が得られる：
\begin{align}
\hat{Y}_{\gamma_1} \hat{Y}_{\gamma_2} = q^{\langle \gamma_1, \gamma_2 \rangle }\hat{Y}_{\gamma_2}\hat{Y}_{\gamma_1} \ .
\label{qTorus}
\end{align}
ただし，$q:=e^{i \hbar}$であり，古典極限は$q\to 1$となる．
このように，4次元理論（従ってM理論）への持ち上げは，ヒッチン・モジュライを量子化するのである\footnote{もっとも，  この主張は最終的には場の理論的による直接計算によって確かめられるべきものである．この方向については例えば\cite{Okuda}を参照． }．

\section{箙の変異とフリップ}

ここまで，M理論を考えることでヒッチン・モジュライが量子化されることを
みてきたが，それでは箙の変異はどう量子化されるのだろうか．

\begin{figure}
\centering\includegraphics[scale=0.4]{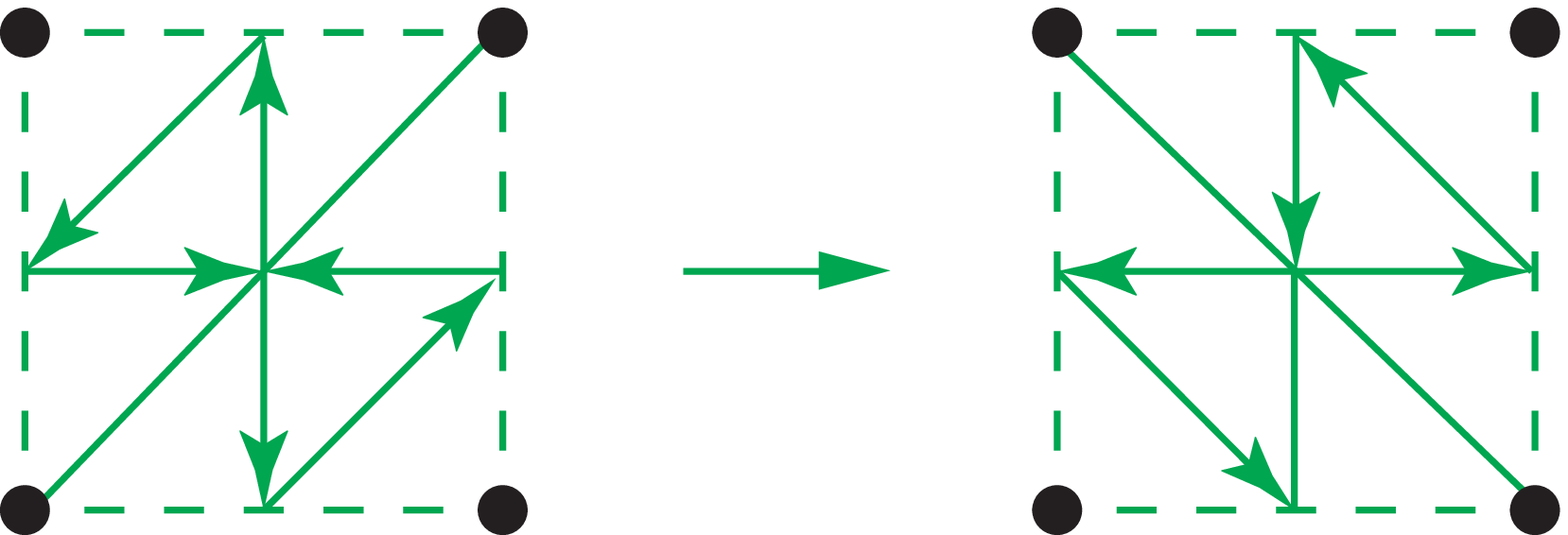}
\caption{理想三角形分割に対するフリップはBPS箙の変異を引き起こす}
\label{trig_mutation}
\end{figure}

ここでは，この変異を組み合わせ論的に説明しておこう．
箙は理想三角形分割から定まっていたのであるから，
箙の曖昧さは理想三角形分割の曖昧さに起因していることになる\footnote{ここでは三角形分割を天下りかつ組み合わせ論的に
与えたが，実際にはより物理的な説明がある：BPS状態はリーマン面$C$上の測地線で与えられ，その測地線の族を考えることで，
理想三角形分割を構成することができる．
また，その三角形分割は$\zeta$の値に依存し，これを変えることで
三角形分割の変化を引き起こすことができる\cite{GMN}．
その構成は（より厳密な設定で）本特集中の岩木氏の記事で取り扱われている．
}．

理想三角形分割の任意性は，理想四角形の対角線を取り替える操作（\textbf{フリップ}と呼ばれる）を繰り返すことによって
尽くされることが知られている．そこで，フリップのもとで変数
$\hat{Y}_{\gamma}$が変化するかを調べればよい．

まず，辺$k$においてフリップをした時，対応する箙$B$は
頂点\eqref{trig_mutation}における変異$\mu_k(B)$に移り変わる（図\ref{trig_mutation}）．
このときチャージは\eqref{gchange_1}及び\eqref{gchange_2}に従って
変化するので，対応する変数$\hat{Y}_{\gamma_k}$も
\begin{align}
\mu_k: 
\begin{array}{l}
\hat{Y}_{\gamma_k} \to \hat{Y}_{-\gamma_k}=\hat{Y}_{\gamma_k}^{-1} \ ,\\
\hat{Y}_{\gamma_i} \to \hat{Y}_{\gamma_i+ [b_{ik}]_{+} \gamma_k} \quad (i\ne k)
\end{array}
\label{Xchange_1}
\end{align}
と変化するのが自然である．
実際，$\hat{Y}_{\gamma_i}$が\eqref{omega}を満たす時，こうして変換した後の$\hat{Y}_{\gamma}$は新しい箙$B'=\mu_k(B)$に対する\eqref{omega}を満たすことが確認できる（なお，このために
\begin{align}
\hat{Y}_{\gamma_i+ [b_{ik}]_{+} \gamma_k}
=q^{-\frac{1}{2}[b_{ik}]_{+} b_{ki}} \hat{Y}_{\gamma_k}^{[b_{ik}]_{+} } \hat{Y}_{\gamma_i}  
\end{align}を用いる．
）．しかしこれで話は終わりではない．
実際には
\begin{align}
K_{\gamma_k}: \hat{Y}_{\gamma_i} \to \Psi_q(\hat{Y}_{\gamma_k})^{-1} \hat{Y}_{\gamma_i} \Psi_q(\hat{Y}_{\gamma_k}) 
\label{Xchange_2}
\end{align}
なる変換をさらに行った合成
\begin{align}
\bar{\mu}_k := K_{\gamma_k} \mu_k
\end{align}
がフリップによって引き起こされる変換なのである．
ここで，$\Psi_q(x)$は\textbf{量子ダイログ関数}\cite{FK}と呼ばれる特殊関数であり，
\begin{align}
\begin{split}
&\Psi_q(qx;q)=(1+q^{1/2} x)^{-1} \Psi_q(x;q) , \\
&\Psi_q(0;q)=1
\end{split}
\end{align}
という関数関係式によって定義される．
より具体的に書くと，$\bar{\mu}_k$の作用は，$\hat{Y}_i:=\hat{Y}_{\gamma_i}$と書くことにして
\begin{align}
\hat{Y}_k \to \hat{Y}_k^{-1}  
\end{align}
及び$i\ne k$の時，$s_{jk}:= \textrm{sgn}(b_{jk} )$と書くと
\begin{align}
\hat{Y}_j \!\to \! 
\prod_{n=0}^{|b_{jk}|-1} \left(1+q^{-(n+\frac{1}{2}) s_{jk}} Y_k^{-s_{jk} }\right)^{-s_{jk}} Y_ j
\end{align}
で与えられる．
これは\textbf{量子クラスター代数}における\textbf{量子$y$変数}の変換則に他ならない．

特に，
$q=1$とすると．
次の変換則が得られる：
\begin{align}
\begin{split}
&Y_k \to Y_k^{-1} \ , \\
&Y_j \to  (1+ Y_k^{-\textrm{sgn}(b_{jk})} )^{-b_{jk}}  Y_j\ .
\end{split}
\end{align}
これはすでにいくつかの記事にも現れた
（古典）クラスター代数の係数（\textbf{$y$変数}）の変換則に他ならない．

ここまでではただ1回の箙の変異を考えたが，
変異を繰り返すこともできる．
その結果箙$B$は別の箙$B'$に変化し，
\begin{align}
B'=\mu_k \cdots \mu_1(B) \ ,
\end{align}
演算子$\bar{\mu}_k$の積は$B$の量子トーラスから$B'$の量子トーラスへの写像
\begin{align}
\bar{\mu}_k\cdots \bar{\mu}_1 
\end{align}
を与える．この形の演算子は，コンツェビッチとソイベルマンによる
\textbf{壁越え現象}における公式の記述に用いられ，
関連して量子ダイログ関数の恒等式やYシステムの構成に用いられる．
また，4次元$\mathcal{N}=2$理論の境界に現れる
3次元$\mathcal{N}=2$理論の分配関数としても
解釈することができる\cite{Terashima:2013fg}．
後者は3次元多様体の幾何にも関係しておりそれ自体
興味深い理論である（本特集寺嶋氏の記事並びに\cite{Yamazaki_book,Yamazaki_Science}を参照）．

\section{団代数の彼方へ}

以上，団代数の構造が超対称ゲージ理論の一つの文脈でどのように現れるかを解説してきた．
団代数とは，4次元$\mathcal{N}=2$理論の\textbf{ループ演算子のなす代数}，及びそれが
チャージ・ラティス$\Gamma_{\rm 正}$の取り換えでどう変化するかを記述するものに他ならないのであった．より一般の超対称ゲージ理論に対しても，ループ演算子のなす演算子を
考えることにより同様の構造が得られると期待される（例えば\cite{KW}を参照）．

興味深いことに，純粋に数学上の興味から生まれた団代数は，超対称場の理論の物理的考察においても重要な構造であるのだ．筆者自身，当初は団代数のことを食わず嫌いでなんとなく敬遠していたが，折に触れて団代数のことは小耳に挟んでいた．その後，自分の研究の中でその強力さに気づき，宗旨替えをして現在に至っている\footnote{筆者の場合，一般的にいって，そうした紆余曲折の後に学んだ知識の方が役立つことが多い．}．この記事をきっかけに，読者の皆さんが
団代数に少しでも親しみを感じて頂ければと願う．

本特集からも明らかなように
団代数は様々な文脈で姿をみせる．超対称場の理論での現れは団代数の
数多くの姿のうちの一つでしかないということもできるだろう\footnote{クラスター代数の数理物理における広がりをのぞくには例えば特集\cite{special}の論文たちを参考されたい．
}．しかし，ここまで議論してきた超対称ゲージ理論の物理は
骨組みとしての代数的な団代数の構造そのものよりもはるかに豊穣な内容を
含んでいる．それは，団代数の多様な側面を一つにまとめてみせると同時に，
新たな数学の発展を刺激してきた．団代数からさらに何を汲み取ることができるのか，その先には何が待ち構えているのか，それを思うとき筆者の胸は高鳴る．

\end{document}